\documentstyle[prc,psfig,aps,twocolumn]{revtex}

\newcommand{\mbf}[1]{\mbox{\boldmath$#1$}}
\begin{document}

\draft

\twocolumn[
\hsize\textwidth\columnwidth\hsize\csname @twocolumnfalse\endcsname

\title{Three-body properties in hot and dense nuclear matter}

\author{M. Beyer$^1$, W. Schadow$^2$, C. Kuhrts$^1$, G. R\"opke$^1$
%}
%\address{
\\
$^1$FB Physik,
 Universit\"at Rostock, Universit\"atsplatz 1, 
18051 Rostock, Germany\\
$^2$
Institute of Nuclear and Particle Physics,  and
Department of Physics,  Ohio University, Athens, OH 45701}

\vspace{10mm}

\date{February 2, 1999}

\maketitle

\begin{abstract}
  We derive three-body equations valid at finite densities and
  temperatures. These are based on the cluster mean field approach
  consistently including proper self energy corrections and the Pauli
  blocking. As an application we investigate the binding energies of
  triton and determine the Mott densities and momenta relevant for a
  many particle description of nuclear matter in a generalized
  Beth-Uhlenbeck approach. The method, however is not restricted to
  nuclear physics problems but may also be relevant, e.g.,  to treat 
  three-particle correlations in weekly doped semiconducters or strongly
  coupled dense plasmas.
\end{abstract}


\vspace{5mm}

]

\section{Introduction}

Correlated many particle systems such as, e.g., nuclear matter or
strongly coupled plasmas, have a complicated dynamical behavior.  Only
few areas of the density-temperature phase diagram used to
characterize the state of the system in thermal equilibrium may be
described in the approximation of noninteracting quasiparticles. The
dynamics of the quasiparticle is determined by the mean field of the
other particles and therefore some prominent features like self energy
corrections and Pauli blocking (Bose enhancement) are sufficiently
taken into account~\cite{fet71}. However due to sizable residual
interactions many interesting and exciting phenomena, such as
clustering, formation of condensates, and phase transitions occur.

A large number of these phenomena such as, e.g., two-particle bound
state formation, can already be accounted for by explicitly
introducing two-body correlations into the formalism. This may be
achieved in the frame work of the Green function method~\cite{fet71}
and leads to effective two-body equations that include medium effects
in a consistent way.

Recently, generic three-body processes have been calculated utilizing
exact few-body methods in the context of many particle systems for i)
nuclear matter~\cite{bey96,bey97} and ii) plasmas at star
conditions~\cite{belyaev}. Although the relevance of three-body
processes in many particle systems have already been recognized,
little progress has been achieved since both fields (many-particle and
few-body physics) in itself are rather elaborated, bearing their own
technical problems. However, some phenomena of many particle systems
{\em require} a treatment of effective few-body systems embedded in a
medium, in particular three-body processes.

In the context of nuclear matter, a three-body Faddeev-type equation
has been derived within the Green functions method applicable to
describe three-body correlations in nuclear matter of thermal
equilibrium~\cite{bey96,bey97}. Empirical evidence, including recent
experimental data on cluster formation~\cite{GSI93,MSU95}, indicate
that a large fraction of deuterons can be formed in heavy-ion
collisions of energies below $E/A\le 200$ MeV. The abundances of
deuterons are determined by the deuteron formation via $NNN\rightarrow
dN$ ($N$ nucleon, $d$ deuteron) and break-up, $dN\rightarrow NNN$,
reactions. The hadronic reaction requires a proper treatment of the
effective three-body problem.  We have numerically solved the
scattering problem to describe the deuteron break-up reaction
consistently including the effects of finite densities and
temperatures of the surrounding nuclear matter.  Within linear
response theory we have given a first estimate for the break-up time
of deuterons embedded in nuclear matter~\cite{bey97}.

Indeed, at moderate densities $n\lesssim n_0/10$ ($n_0=0.17$~fm$^3$
nuclear matter density) and temperatures $T\lesssim 15$~MeV nuclear
matter may be considered as a mixture of light nuclei (chemical
picture). The transition into that region is determined by the Mott
density~\cite{roe83} and is relevant, e.g. for the final stage of a
heavy ion collision at intermediate energies, respectively on the
surface of the expanding nuclear matter. The Mott effect has also
been considered at the beginning of the heavy ion reaction using
inverse photo disintegration~\cite{boz98}. Here we present a
calculation that determine the Mott densities for the three-body
nuclear bound state.

%=================================================================
\section{Beth-Uhlenbeck approach to nuclear matter}
\begin{figure}[p]
  \leavevmode
\centering
 \psfig{figure=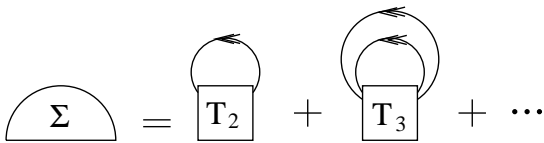,width=0.4\textwidth}
\vspace*{1ex}
\caption{\label{fig:sigcl} Cluster decomposition of the self energy
  $\Sigma$ in terms of $n$-body $t$-matrices. The sum is over
  irreducible contributions only.} 
\end{figure}

%=================================================================
The basis to treat correlated densities is
provided by a generalization of the Beth-Uhlenbeck
approach~\cite{sch90}. The nuclear density $n=n(\mu,T)$ as a function
of the chemical potential $\mu$ (for the time being we assume
symmetric nuclear matter) and temperature $T$ may be written as
\begin{equation}
n=n_{\rm free} + n_{\rm corr},\qquad 
n_{\rm free} = 4\sum_1 f_1.
\end{equation}
For abbreviation we use $1\equiv\alpha_1=\{k_1, s_1, \tau_1, \dots\}$
denoting momenta, spin, isospin, etc. of particle 1.  The one-particle
Fermi function is 
\begin{equation}
f_1\equiv f(\varepsilon_1)
  =(\exp[\beta(\varepsilon_1-\mu)]+1)^{-1},
\end{equation}
where $\varepsilon_1,$ denotes the quasiparticle energy and $\beta$
the inverse temperature.  In first iteration the correlations may be
treated on the basis of residual interactions between the
quasiparticles.  To do so the imaginary part of the self energy
$\Sigma (1,\omega)$ should be small and the respective spectral
function may be expanded with respect to the imaginary part $\Sigma_I
(1,\omega)$ as explained in Ref.~\cite{sch90}. The contribution of the
correlated density to the total density may then be written as
\cite{sch90}
\begin{eqnarray}\nonumber
n_{\rm corr}(\mu,T)&=&\sum_1 \int \frac{d\omega}{2\pi} \Sigma_I
(1,\omega-i0^+)\\
\times& & [f(\omega)-f(\varepsilon_1)]\frac{d}{d\omega}\frac{{\cal P}}
{\varepsilon_1-\omega},\label{eqn3}
\end{eqnarray}
where ${\cal P}$ denotes principle part integration. To evaluate the
self energy we utilize cluster decomposition taking into account two-
and three-particle, in general n-particle correlations (see
Fig.~\ref{fig:sigcl}),
\begin{eqnarray}\nonumber
\Sigma (1,z_\nu)&=&\sum_2 \sum_{\lambda} T_2
(12,12;z_{\nu}+z_{\lambda}) G_1(2,z_{\lambda})\\\nonumber
&+&\sum_{2,3} \sum_{\lambda \lambda'} T^c_3
(123,123;z_{\nu}+z_{\lambda}+z_{\lambda'})\\\nonumber
&& \qquad\qquad \times  G_1(2,z_{\lambda})G_1(3,z_{\lambda'})\\
&+& \dots\label{eqnSig}
\end{eqnarray}
where $G_1(z_{\nu })$ denotes the one-particle Green function of
Matsubara frequency $z_{\nu}$ \cite{fet71}, and the index $c$ denotes
connected terms only. The $n$-body $t$-matrices are defined in terms
of the $n$-body Green functions $G_n(z)$ via
\begin{eqnarray}
G_n (z)=G_n^{(0)}(z)+G_n^{(0)}(z)T_n (z)G_n^{(0)}(z).
\end{eqnarray}
Using the spectral representation of the $t$-matrices and evaluating
the Matsubara sums eventually leads to the following decomposition of
the correlated density induced by the cluster decompostion Eq. 
(\ref{eqnSig}),
\begin{equation}
n_{\rm corr.} = 2n_2+3n_3 +\dots, \qquad
n_2=n_2^{\rm b} + n_2^{\rm sc},\dots
\label{eqn:corr}
\end{equation}
where $n_{2(3)}$ denotes the two- (three-) particle correlated
densities, present as bound $n^{\rm b}$ or scattering $n^{\rm sc}$
states in chemical equilibrium.  For the two-particle case the
correlated densities have been given explicitly in Ref.~\cite{sch90}
\begin{eqnarray}
n_2^{\rm b} &=&\sum_{P>P_{\rm Mott}} 3 
\;g(E_{\rm cont}+E_{\rm b}),\nonumber\\
n_2^{\rm sc} &=& -\sum_{P>P_{\rm Mott}} 3\; g(E_{\rm cont})\nonumber\\
&&-\sum_P \int \frac{d\omega}{2\pi}
\left[\frac{d}{d\omega} g(E_{\rm cont}+\omega)\right]
\nonumber\\
&&\times\sum_\alpha C_\alpha\cdot(\delta_\alpha - 
\sin\delta_\alpha\cos\delta_\alpha),
\label{eqn:bsc}
\end{eqnarray}

where the sum over $\alpha$ indicates partial wave decomposition with
$C_\alpha$ the proper Clebsch-Gordan coefficient and $\delta_\alpha$
the corresponding phase shift. The energies appearing in
Eq.~(\ref{eqn:bsc}) are the binding energy $E_{\rm b}$ and the
continuum energy $E_{\rm cont}$ that includes the proper self energy
shifts. The Bose function for the two-fermion system is
\begin{equation}
g(\omega) = (e^{\beta (\omega-2\mu)} -1)^{-1}.
\end{equation} 
To evaluate the contribution of the three-body correlated density
$n_3$ we presently focus on the bound state contribution. The
$t$-matrix is then dominated by the bound state pole of energy $E_{\rm
  b}$.  Evaluating the respective Matsubara sums for the three-body
bound state contribution in Eq.~(\ref{eqn3}) using
Eq.~(\ref{eqnSig})and the spectral representation of $T_3$ leads to a
rather simple expression for the three-body bound state contribution
$n^{\rm b}_3$ to the total density,viz.
\begin{equation}
n_3^{\rm b} =\sum_{P>P_{\rm Mott}} 4 \;
(e^{\beta (E_{\rm cont}+E_{\rm b}-3\mu)} +1)^{-1},
\end{equation}
where $E_{\rm cont}$ the respective continuum energy.

At moderate densities not only two-particle but also three- and
four-particle correlations occur. Here we consider the three-nucleon
bound state within the generalized Faddeev approach discussed in the
next section. One obvious feature in the above equation is the
appearing of the Mott momentum $P_{\rm Mott}$. It is important in the
framework of the cluster Hartree-Fock expansion~\cite{roe83,chf}
closely related to the self consistent RPA~\cite{scrpa} and extended
to finite temperature in \cite{bey96,bey97,duk98} that are followed
here to consistently treat few-body correlations in matter.

%=================================================================
\section{Finite temperature three-body equations -- bound states}
\label{sec:green}
%=================================================================

The formalism to derive few-body Green functions within the cluster
mean field approximation at finite temperatures and densities has been
given elsewhere~\cite{bey97,duk98}. Here we give some of the basic
results and extend the formalism to include bound states. The reaction
cross section based on the suitably modified AGS formalism has been
given in~\cite{bey96} and the life time of deuteron fluctuations in
nuclear matter as a first application in~\cite{bey97}.
  
The hierarchy of Green function equations is truncated using cluster
mean field expansion, and a calculable form is achieved by introducing
ladder approximation. We consider generic elementary two-particle
interactions $V_2$ only. The resulting equations for the decoupled
one-, two- and three-body Green functions at finite
temperatures (utilizing the Matsubara technique to treat finite
temperatures) will be given in the following. The one-particle Green
function reads
\begin{equation}
G_1(z) = R_1^{(0)}(z) = \left(z - \varepsilon_1\right)^{-1}.
\label{eqn:G1}
\end{equation}
In mean field approximation the single quasiparticle
energy $\varepsilon_1$ is given by
\begin{eqnarray}
\varepsilon_1 &= &\frac{k^2_1}{2m_1} + \Sigma^{HF}(1),\nonumber \\
\Sigma^{HF}(1) &=& \sum_{2}
\left[ V_2(12,12)-  V_2(12,21) \right]\,  f_2,
\label{eqn:selfHF}
\end{eqnarray}
and $f_2=f(\epsilon_2)$ the Fermi function given in the previous
section.  The equation for the two-particle Green function $G_2(z)$
reads
\begin{equation}
G_2(z) = N_2R_2^{(0)}(z) + R_2^{(0)}(z)N_2V_2 G_2(z),
\label{eqn:G2}
\end{equation}
where the two-body resolvent $R_2^{(0)}(z)$ is given by
\begin{equation}
 R_2^{(0)}(12,1'2';z) = \frac{\delta_{11'}\delta_{22'}}
{z - \varepsilon_1 - \varepsilon_2}
\end{equation}
and the Pauli blocking factor $N_2$ by
\begin{eqnarray}
N_2(12,1'2') &= &\delta_{11'}\delta_{22'}\delta_{33'}
(\bar f_1 \bar f_2 - f_1 f_2)\nonumber\\ 
&=&\delta_{11'}\delta_{22'}\delta_{33'}
(1 - f_1 -  f_2).
\end{eqnarray}
 We use the notation $\bar f = 1-f$.
The respective equation for the three-particle Green function relevant
to describe three-body correlations in a medium is given by
\begin{equation}
G_3(z) = N_3R_3^{(0)}(z) + R_3^{(0)}(z) W_3 G_3(z),
\label{eqn:G3}
\end{equation}
where the effective potential $W_3$ reads
\begin{eqnarray}
  W_3(123,1'2'3') &=&\sum_{k=1}^3 W^{(k)}_3(123,1'2'3'),
\label{eqn:Vchn}\\
W^{(3)}_3(123,1'2'3')&=& (1-f_1-f_2) V_2(12,1'2')\delta_{33'}.
\label{eqn:V3}
\end{eqnarray}
The last Eq. is given for $k=3$ and cyclic permutation is understood.
Note that $W_3\neq W_3^\dagger$. The Pauli factors $N_3$ and the
resolvents $R_3^{(0)}$ read respectively
\begin{eqnarray}
N_3(123,1'2'3') &= &\delta_{11'}\delta_{22'}\delta_{33'}
(\bar f_1 \bar f_2 \bar f_3 + f_1 f_2  f_3)\label{eqn:FPauli}\\
 R_3^{(0)}(123,1'2'3';z) &= &\frac{\delta_{11'}\delta_{22'}\delta_{33'}}
{z - \varepsilon_1 - \varepsilon_2 - \varepsilon_3}.
\end{eqnarray}
Note that $[N_3,R_3^{(0)}]=0$.  
For convenience we may
introduce the Green function of the noninteracting system, viz.
\begin{equation}
G_3^{(0)}(z)=N_3R_3^{(0)}(z).
\label{eqn:G0}
\end{equation}
If we now introduce a potential
$V_3 = N_3^{-1}W_3$ we may instead of Eq.~(\ref{eqn:G3}) write
\begin{equation}
G_3(z) = G_3^{(0)}(z) + G_3^{(0)}(z) V_3 G_3(z),
\label{eqn:G3neu}
\end{equation}
which looks formally as the equation for the isolated
case~\cite{glo88} and allows one to use three-body techniques to
arrive at a solvable form, e.g. AGS equations for the transition
operator\cite{bey97,alt67}.  These will be used to derive numerically
solvable Faddeev type equations at finite temperature.  Although we
assume a Fermionic system the proper symmetrization is treated
separately.  
%\begin{equation}
%V^{(3)}_3(123,1'2'3')= (1-f_3+g(\varepsilon_1+\varepsilon_2))^{-1}
%V_2(12,1'2')\delta_{33'},
%\label{eqn:V3neu}
%\end{equation}
 
Already in Eq.~(\ref{eqn:Vchn}) we have introduced the channel
notation that is convenient to treat systems with more than two
particles~\cite{glo88}. In the three-particle system usually the
index of the spectator particle is used to characterize the channel.

If the correlated pair, e.g. (12) and the spectator particle, e.g. 3,
are uncorrelated in the channel $(3)$ we may define a channel Green
function $G_3^{(3)}(z)$. We generalize to the channel $(\gamma)$. The 
channel Green function is then defined by
\begin{equation}
G_3^{(\gamma)}(z) = \frac{1}{-i\beta} \sum_\lambda\;
iG_2(\omega_\lambda)\;G_1(z-\omega_\lambda). 
\label{eqn:Gchanneldef}
\end{equation}
The summation is done over the Bosonic Matsubara frequencies
$\omega_\lambda$, $\lambda$ even, $\omega_\lambda=\pi\lambda/(-i\beta) +2\mu$.  The equation
for the channel Green function is derived in the same way as for the
total three-particle Green function given in Eqs.~(\ref{eqn:G3}) and
(\ref{eqn:G3neu}). The result is
\begin{equation}
G_3^{(\gamma)}(z) = G_3^{(0)}(z) + G_3^{(0)}(z)
  V^{(\gamma)}_3 G_3^{(\gamma)}(z),
\label{eqn:Gchannel}
\end{equation}
(no summation of $\gamma$). Introducing the notation $\bar V_3^{(\gamma)} =
V_3 - V_3^{(\gamma)}$ we arrive at the following equation for
$G_3(z)$ expressed through the channel Green functions
$G_3^{(\gamma)}(z)$, i.e.
\begin{equation}
G_3(z) = G_3^{(\gamma)}(z)
+G_3^{(\gamma)}(z) \bar V_3^{(\gamma)} G_3(z).
\label{eqn:G3Ggam}
\end{equation}

Now we have set the necessary equations, i.e. Eqs.~(\ref{eqn:G3neu}),
(\ref{eqn:Gchannel}) and (\ref{eqn:G3Ggam}) to derive a proper
integral equation for scattering (see Ref.~\cite{bey96,bey97}) and
bound states at finite temperatures and densities.

To arrive at a homogeneous three-body equation for the bound state
$|\Psi_t\rangle$ in medium we insert Eq.~(\ref{eqn:G3neu}) into the
Lippmann-Schwinger equation
\begin{equation}
|\Psi_t\rangle=\lim_{\epsilon\rightarrow 0} \; i\epsilon\;
G_3(E_t+i\epsilon)\,|\Psi_t\rangle.
\label{eqn:LS}
\end{equation}
After performing the limit we find
\begin{equation}
|\Psi_t\rangle=G_3^{(0)}(E_t)V_3\,|\Psi_t\rangle.
\end{equation}
The Faddeev components are given by
\begin{equation}
|\Psi^{(\alpha)}\rangle = G_3^{(0)}\,V_3^{(\alpha)}\,|\Psi_t\rangle
\end{equation}
and 
\begin{equation}
|\Psi_t\rangle=\sum_\alpha |\Psi^{(\alpha)}\rangle.
\end{equation}
To arrive at a conveniently solvable version of the three-body bound
state problem in medium we introduce form factors. To this end
Eq.~(\ref{eqn:G3Ggam}) is inserted into Eq.~(\ref{eqn:LS}). The limes
$\epsilon\rightarrow 0$ results in
\begin{equation}
|\Psi_t\rangle=G_3^{(\alpha)}(E_t)\bar V^{(\alpha)}_3\,|\Psi_t\rangle.
\end{equation}
Introducing the usual form factors
\begin{equation}
|F^{(\alpha)}\rangle=\bar V^{(\alpha)}_3\,|\Psi_t\rangle
\end{equation}
eventually leads to an integral equation
\begin{equation}
|F^{(\alpha)}\rangle=\sum_\beta (1-\delta_{\alpha\beta})\,T^{(\beta)}_3\,G^{(0)}_3
|F^{(\beta)}\rangle
\label{eqn:form}
\end{equation}
and finally to the bound state given in terms of the form factors
\begin{equation}
|\Psi_t\rangle = \sum_\beta \, G^{(0)}_3T^{(\beta)}_3\,G^{(0)}_3
\,|F^{(\beta)}\rangle.
\label{eqn:wfform}
\end{equation}
The transition channel operator $
T^{(\beta)}_3$ is defined via
\begin{equation}
G_3^{(\beta)}= G_3^{(0)}+ 
G_3^{(0)}  T^{(\beta)}_3 G_3^{(0)},
\end{equation}
inserting this equation into Eq.~(\ref{eqn:Gchannel})
leads to an equation for the channel $t$ matrix
\begin{equation}
  T^{(\beta)}_3 =   V_3^{(\beta)} + 
G_3^{(0)}  V_3^{(\beta)}   T_3^{(\beta)},
\label{eqn:Tchannel}
\end{equation}
and to $V_3^{(\beta)} G_3^{(\beta)} = T_3^{(\beta)} G_3^{(0)}$.  Writing the
Pauli factors explicitly in the integral
equation~(\ref{eqn:form}) results in
\begin{equation}
|F^{(\alpha)}\rangle=\sum_\beta 
(1-\delta_{\alpha\beta})\,T^{(\beta)}_3\,N_3 R^{(0)}_3
|F^{(\beta)}\rangle.
\end{equation}
Note that $ G^{(0)}_3=N_3 R^{(0)}_3=R^{(0)}_3N_3$. We may now
introduce $T_3^{*(\beta)} =N_3^{1/2} T_3^{(\beta)} N_3^{1/2}$ and
$|F^{*(\beta)}\rangle=N_3^{1/2}|F^{(\beta)}\rangle$. The resulting
equation is
\begin{equation}
|F^{*(\alpha)}\rangle=\sum_\beta 
(1-\delta_{\alpha\beta})\,T^{*(\beta)}_3\, R^{(0)}_3
|F^{*(\beta)}\rangle.
\label{eqn:Fstar}
\end{equation}
The equation for the transition channel operator $T_3^{*(\beta)}$ 
 is then ($V_3^{*(\beta)} =N_3^{1/2} V_3^{(\beta)} N_3^{1/2}$)
\begin{equation}
T_3^{*(\beta)}= V_3^{*(\beta)}+V_3^{*(\beta)}R_3^{(0)}T_3^{*(\beta)}.
\label{eqn:Tchstar}
\end{equation}
Inserting all definitions
the explicit form of the effective potential arising in this equation
reads
%\begin{eqnarray}
%\lefteqn{V_3^{*(3)}(123,1'2'3')
%=(1-f_1-f_2)^{1/2}(1-f_3+g(\varepsilon_1+\varepsilon_2))^{-1/2}} \nonumber\\
%&&\times V_2(12,1'2')\delta_{33'}
%(1-f_3+g(\varepsilon_{1'}+\varepsilon_{2'}))^{1/2}(1-f_{1'}-f_{2'})^{1/2}\\
%&\simeq&(1-f_1-f_2)^{1/2} V_2(12,1'2')(1-f_{1'}-f_{2'})^{1/2}.
%\label{eqn:hermitian}
%\end{eqnarray}
\begin{eqnarray}
\lefteqn{V_3^{*(3)}(123,1'2'3')
=N_2^{1/2}(12)(1-f_3+g(\varepsilon_1+\varepsilon_2))^{-1/2}} \nonumber\\
&&\times V_2(12,1'2')\delta_{33'}
(1-f_3+g(\varepsilon_{1'}+\varepsilon_{2'}))^{1/2}N_2^{1/2}(1'2') \nonumber \\
&\simeq&(1-f_1-f_2)^{1/2} V_2(12,1'2')(1-f_{1'}-f_{2'})^{1/2}.
\label{eqn:hermitian}
\end{eqnarray}
where we have used $(f_1f_2f_3 + \bar f_1 \bar f_2 \bar
f_3)=(1-f_i-f_j)(1-f_3+g(\varepsilon_1+\varepsilon_2))$ valid for all
permutations of $ijk=123$. Note that $1-f_1-f_2>0$ which is the case
for low densities and the last equality in Eq.~(\ref{eqn:hermitian})
holds for $f^2\ll f$.  Utilizing this approximation the corresponding
scattering solution using a separable ansatz for the strong
nucleon-nucleon potential has been given in~\cite{bey96}. In
Ref.~\cite{bey96} we have calculated the break up cross section
$Nd\rightarrow NNN$ and found considerable dependence of the deuteron
fluctuation time on the proper treatment of the medium
dependence~\cite{bey97}.  Here we solve Eq.~(\ref{eqn:Fstar}) for the
Yamaguchi~\cite{yamaguchi} and a high rank separable version (Paris
(EST)) of the Paris potential~\cite{paris}.  For a detailed overview
on the procedure we refer to \cite{schadow}.  To compared to the
perturbation theory result we use the respective wave function of the
isolated triton and evaluate the medium dependent part, i.e. the Pauli
blocking $-(f_1+f_2)V_2$ and self energy corrections $\Sigma^{HF}(1)$
of the effective Hamiltonian in the standard fashion.

\section{Kinematics}

Unlike the isolated three-body problem Galilei invariance (for the
three-body system embedded in a medium) is not satisfied, since the
Fermi functions depend explicitly on the relative momentum of the
three-nucleon system with respect to the medium. 

\begin{figure}[p]
  \leavevmode
\centering
 \psfig{figure=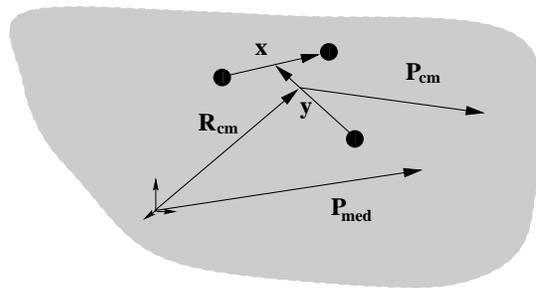,width=0.4\textwidth}
\vspace*{1ex}
\caption{\label{fig:kin} Kinematical variables of the three-body
  system embedded   in a medium.} 
\end{figure}

As a consequence one has to solve the three-body problem at a finite
center of mass momentum $P_{\rm cm}=k_1+k_2+k_3$ in a medium that may
be considered at rest $P_{\rm med}=0$ for simplicity, see
Fig.~\ref{fig:kin}.  However, technically it is more convenient to let
the three-nucleon system rest $P_{\rm cm}=0$ and the surrounding
medium move with $P_{\rm med} = - (k_1+k_2+k_3)$. This procedure
results in the least change of the three-body algebra and is possible
since the dependence on $P_{\rm cm}-P_{\rm med}$ is only through the
Pauli blocking factors thus parametric.

For simplicity we use angle averaged Fermi function $\langle\langle
N_2\rangle\rangle$ and $\langle\langle N_3\rangle\rangle$, i.e.
\begin{equation} 
\langle\langle N_3\rangle\rangle =
\frac{1}{(4\pi)^2}\;\int d\cos\theta_q \; d\cos\theta_p \; d\phi_q
d\phi_p\; N_3(\mbf{p},\mbf{q},\mbf{P}_{\rm cm}), 
\end{equation}
%\begin{equation} 
%\langle\langle N_2\rangle\rangle =
%2\pi\;\int dx\; N_2(\bf{p},\bf{q}), 
%\end{equation}
where the angles are taken with respect to $\mbf{P}_{\rm cm}$ and
$\mbf{p}$, $\mbf{q}$ are the standard Jacobi coordinates~\cite{glo88}.

\section{Results}

Since the Green functions have been evaluated in an independent
particle basis the one-, two-, and three-particle Green functions are
decoupled in hierarchy, as given in Eqs.~(\ref{eqn:G1}),
(\ref{eqn:G2}), (\ref{eqn:G3}). To solve the in medium problem up to
three-particle clusters the one-, two- and three-particle problems are
consistently solved. This leads to the single particle self energy
shift Eq.~(\ref{eqn:selfHF}), the two-body input including the proper
Pauli blocking Eq.~(\ref{eqn:Tchstar}), and eventually to the
three-body bound state Eqs.~(\ref{eqn:wfform}) and (\ref{eqn:Fstar})
(or scattering state).

For technical reasons we have approximated the nucleon self energy
calculated via Eq.~(\ref{eqn:selfHF}) by use of effective masses for the
nucleon, which is reasonable for the small nuclear densities
considered here.

For comparison we have calculated the energy shift of the three-body
bound state using perturbation theory and the solutions of the
respective isolated system.
 
\begin{figure}[p]
  \leavevmode
\centering
 \psfig{figure=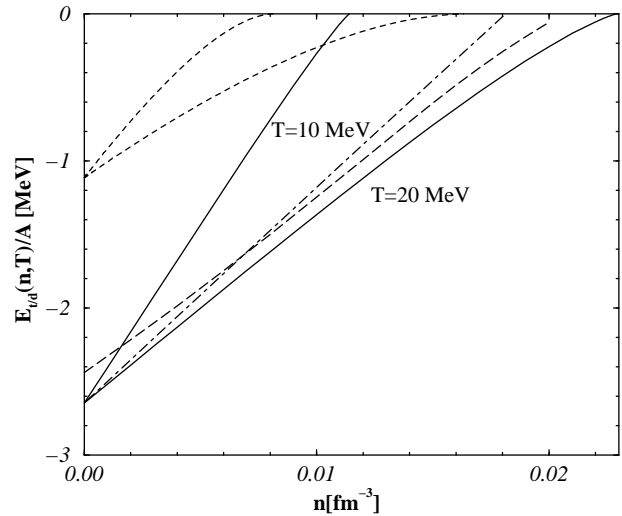,width=0.45\textwidth,angle=270}
\vspace*{1ex}
\caption{\label{fig:tri0} Triton binding energy per nucleon (solid
  lines, Yamaguchi potential) as a function of the uncorrelated
  nuclear density at a given temperature $T$; long-dashed lines:
  corresponding pertubation result, dashed-dotted lines: Paris (EST)
  potential.  Dashed lines show the $Nd$ continuum threshold for
  $T=10$ MeV (left) and $T=20$ MeV (right).}
\end{figure}

The resulting binding energies per nucleon in case where triton and
medium are both at rest, i.e. $P_{\rm cm}=P_{\rm med}=0$, as a
function of the nuclear uncorrelated density $n$ are shown in
Fig.~\ref{fig:tri0} for two different temperatures $T=10$, $20$
MeV. The rank one Yamaguchi potential has been used for the solid
lines, also used earlier in the context of in medium break up
reactions ($Nd\rightarrow NNN$)~\cite{bey96}. The dashed line
originating at the deuteron binding energy $E_d/2\simeq -1.11$ MeV
reflects the respective continuum threshold (since the deuteron also
changes the binding energy the continuum threshold changes). The
intersection between the triton binding energy and the continuum
defines the Mott transition density. The Mott transition of the triton
leads for both temperatures considered directly to a three-body
break-up.

For $T=20$ MeV the long dashed line shows the result of the Paris
(EST)  potential. The difference between the dashed line and the
respective result of the simple Yamaguchi potential is mostly due to
the difference in binding energy.  The shape of the curves, also for
the case $T=10$ and $30$ MeV (not shown) are very similar. The dashed
dotted line is the result of a perturbative calculation of all medium
effects evaluated by using the triton wave functions for the isolated
case. The dominant effect comes from Pauli blocking.  The self energy
correction including the continuum shift (due to effective masses) is
smaller.

\begin{figure}[p]
  \leavevmode
\centering
  \psfig{figure=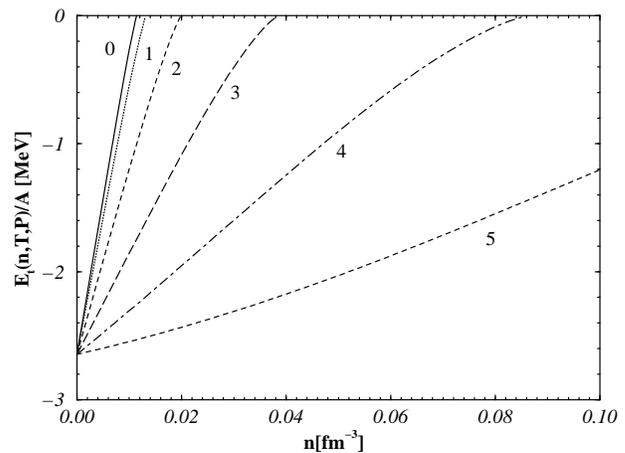,width=0.45\textwidth,angle=270}
\vspace*{1ex}
\caption{\label{fig:triP} Triton binding energy as a function of
  nuclear density at $T=10$ MeV, and total momentum relative to the
  medium. From left to right:
  $P=0,1,2,3,4,5$ fm$^{-1}$.}
\end{figure}

The momentum dependence of the binding energy per nucleon for $T=10$
MeV is shown in Fig.~\ref{fig:triP}. As expected and known from the
deuteron case the influence of the surrounding matter on the binding
energy is decreasing with increasing momentum. Therefore the Mott
transition moves to higher densities for higher momenta. As a
consequence in a moving system (like in a heavy ion collision)
deuterons, tritons and presumably other light clusters may be formed
at higher densities than expected from the simple considerations at
rest. The momentum where the triton binding energy crosses the
continuum is referred to as Mott momentum $P_{\rm Mott}$.

\begin{figure}[p]
  \leavevmode
\centering
 \psfig{figure=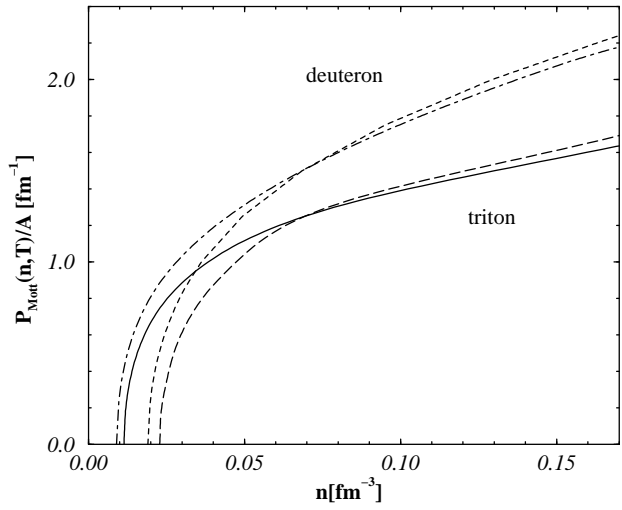,width=0.45\textwidth,angle=270}
\vspace*{1ex}
\caption{\label{fig:mottP} Mott momentum per nucleon $P_{\rm Mott}/A$
  for triton - solid ($T=10$ MeV) and dashed line ($T=20$ MeV) and
  deuteron - dashed dotted ($T=10$ MeV) and short dashed line ($T=20$
  MeV) as a function of nuclear density. Below the respective lines
  tritons (deuterons) do not exist as bound states.}
\end{figure}

Finally, Fig.~\ref{fig:mottP} shows the dependence of the Mott
momentum on the density for the triton and the deuteron. The momenta
and the densities related to the area above the respective curves
allow bound states to be formed (i.e. $E_{t,d}<0$). At higher
densities (respectively at higher momenta) the temperature dependence
is less pronounced than for the lower densities.

\section{Summary and conclusions}

We have given proper equations to treat three-body correlations in
matter and solved them for the nuclear bound state. The framework is
provided by the cluster mean field (cluster Hartree-Fock)
approximation closely related to the self consistent RPA approach,
known for $T=0$, and generalized to $T\neq 0$
recently~\cite{bey97,duk98}.  

We find substantial changes of the triton properties (being a much
simpler problem the deuteron has been discussed much
earlier~\cite{roe83}) and give  the respective Mott densities and
momenta. 

The application of rigorous few-body methods in the context of a
many-particle description of nuclear matter provides a fruitful method
not only useful for nuclear physics. Potential areas for a new
application may be, e.g., weekly doped semiconductors to include the
effects of bound $eeh$ charged exciton states (trions) or strongly
coupled dense plasmas where ionisation rates may rely on few-body
reactions if the charge of the ions become large.

The method is capable to be applied to the $\alpha$ particle as
well, which is the strongest bound nucleus and therefore very
important for nuclear matter. Recently, indication for
$\alpha$-particle quartetting (condensation) has been found in a
rather simple approach that need to be confirmed~\cite{alpha}.

\section{Acknowledgment}
This work has been supported by the Deutsche Forschungsgemeinschaft.
The work of W.S. was performed under the auspices of the U.~S.
Department of Energy under contract No. DE-FG02-93ER40756 with Ohio
University. 
\appendix

%  \begin{thebibliography}{99}


\end{document}